\documentclass[final,5p,times,twocolumn]{elsarticle}

\usepackage{amssymb}
\usepackage{lipsum}
\usepackage{amsthm}
\usepackage{mathtools}
\usepackage{siunitx}
\usepackage{xcolor}
\usepackage{graphicx}
\usepackage{adjustbox}
\usepackage{placeins}
\usepackage[T1]{fontenc}
\usepackage{lipsum}
\usepackage{csquotes}
\usepackage{enumerate}
\usepackage{hyperref}
\usepackage{siunitx}

\newcommand{\citerange}[2]{\cite{#1}--\cite{#2}}

\journal{Journal of High Energy Astrophysics}

\begin{document}

\begin{frontmatter}

\title{Charge constraint on M87* with twisted light}

\author[first,second,third]{Fabiano Feleppa}
\affiliation[first]{organization={Dipartimento di Fisica ``E.R. Caianiello'', Università di Salerno},
addressline={Via Giovanni Paolo II 132}, 
city={I-84084 Fisciano},
country={Italy}}
\affiliation[second]{organization={Istituto Nazionale di Fisica Nucleare, Sezione di Napoli},
addressline={Via Cintia, 80126}, 
city={Napoli},
country={Italy}}
\affiliation[third]{organization={Nordic Institute for Theoretical Physics (NORDITA)},
addressline={Hannes Alfvéns väg 12}, 
city={SE-114 19 Stockholm},
country={Sweden}}

\author[fourth,fifth]{Fabrizio Tamburini}
\affiliation[fourth]{organization={Rotonium Quantum Computing},
addressline={Le Village by CA, Pz.\ G.\ Zanellato, 23}, 
city={I-35131 Padova},
country={Italy}}
\affiliation[fifth]{organization={INAF - Istituto Nazionale di Astrofisica},
addressline={Vicolo dell’Osservatorio 5}, 
city={I-35122 Padova},
country={Italy}}

\author[first,second]{Gaetano Lambiase}

\begin{abstract}
We propose a novel method to constrain the electric charge of the supermassive black hole M87* by analyzing the orbital angular momentum content of the light it emits. By leveraging the established analogy between rotating spacetimes and inhomogeneous optical media, we derive a simple analytical formula that relates the average orbital angular momentum in the observed radiation to the black hole’s charge-to-mass ratio. Applying this relation to existing observational data, we place an upper bound of $\mathcal{Q}/M \lesssim 0.39$ on the charge of M87*. While the analysis focuses on electric charge, which is used here purely as a theoretical example since astrophysical black holes are expected to be approximately neutral, the method is general and can be extended to constrain other types of charges -- degrees of freedom that define distinct black hole solutions. These results demonstrate the potential of orbital angular momentum as a new fundamental degree of freedom to be exploited in astrophysics, providing a complementary and independent alternative to shadow-based techniques for probing the properties of rotating compact objects and testing gravity in the strong-field regime.
\end{abstract}

\begin{keyword}
supermassive black holes \sep twisted light \sep spin-to-orbital angular momentum conversion

\end{keyword}

\end{frontmatter}

General relativity provides a consistent framework for describing the interaction between gravitational fields and matter through the curvature of spacetime. Predictions derived from this theory have been extensively tested and validated in the weak-field regime through a wide range of high-precision solar-system experiments \cite{Will2014,Collett2018}. The success of general relativity has also been confirmed in the strong-field regime, through observations of binary pulsars \cite{Damour1992,Wex2020} and measurements of gravitational redshift \cite{Abuter2018} and geodetic orbit-precession \cite{Abuter2020} of the star S2 near our galaxy’s central supermassive compact object, Sgr A*, by the GRAVITY collaboration. These results confirm general relativity's status as the most well-established classical theory of gravity to date. 

A few years ago, a significant step forward has been made in testing general relativity at the strongest field strengths, thanks to the development of Very Long Baseline Interferometry (VLBI). This advancement has enabled the capture of horizon-scale images of supermassive black holes. As a result, a new era has been ushered in, providing new opportunities to further test fundamental physics. In particular, a major milestone in this field was achieved using the Event Horizon Telescope (EHT), a millimeter VLBI array with Earth-scale baselines \cite{Fish2016}. In 2019, the EHT collaboration produced the first resolved images of the near-horizon region of M87* \citerange{L1}{L6} (see also Ref.\ \cite{Agol2000}, which presents the original idea underlying such observations). Subsequent observations revealed its magnetic field structure \cite{Polarization1,Polarization2}, further characterizing the black hole's phenomenology. More recently, the first images of Sgr A* were obtained from EHT data, providing another crucial dataset for testing gravity and fundamental physics \citerange{L12}{L17}. These horizon-scale images reveal key features of black holes, most notably a bright emission encircling a central brightness depression, the latter being known as the `shadow' of the black hole (for detailed reviews on the black hole shadow and related concepts, see, e.g., Refs.\ \citerange{Bozza2010}{Chen2023}).

Current EHT observations do not yet allow to rule out theories alternative to general relativity. However, recent analyses have used the shadow size of M87* to place constraints on the physical charges — parameters that characterize different black hole solutions — across a wide range of rotating and nonrotating black holes \cite{Kocherlakota2021,Vagnozzi2023}. Notably, these constraints have already excluded the possibility that M87* is a highly charged dilaton black hole, for example. In Ref.\ \cite{Kocherlakota2021}, a constraint on the electric charge of M87* is also derived, under the assumption that it is described by the Reissner-Nordström solution. By analyzing the deviation of the shadow size of a Reissner-Nordström black hole from that of a Schwarzschild black hole, the authors establish an upper limit on the charge-to-mass ratio, $\mathcal{Q}/M \lesssim 0.90$, at the 1$\sigma$ confidence level. It is also noteworthy that a slightly less stringent constraint ($\mathcal{Q}/M \lesssim 0.96$) was recently obtained in Ref.\ \cite{Tsukamoto2024}, where the authors adopted different assumptions compared to those used in Ref.\ \cite{Kocherlakota2021}. For additional studies on constraints on black hole charges — arising from both theoretical arguments and observational data, and not necessarily limited to electrical charge — the reader is referred to, for example, Refs.\ \citerange{Zakharov2014}{Pantig2023}.

In this work, our aim is to obtain a constraint on the electric charge of M87* by exploiting the information related to the orbital angular momentum (OAM) of the radiation emitted by the black hole. In particular, we will assume that M87* is described by the Kerr-Newman solution \cite{Newman1965}.\ Although astrophysical black holes are not expected to possess an electric charge due to neutralization via accreting plasma \cite{Wald1974,Blandford1977}, we consider this example solely for simplicity. The analysis presented here is not intended to suggest that M87* is actually charged; rather, this work is a proof-of-concept aimed at showing how OAM can be exploited to constrain different geometries. Our approach complements shadow-based techniques and can, in principle, be extended to other types of charges or modifications to general relativity.

First theorized by Abraham in 1914 \cite{Abraham1914}, OAM is one of the two components of total electromagnetic angular momentum, the other being spin angular momentum (SAM). Over the years, OAM has found a broad range of applications across multiple fields, including nanotechnology \cite{Grier2003,Shen2019}, quantum cryptography \cite{Zeilinger2002,Padgett2015} and telecommunications \citerange{Padgett2004}{Spinello2016}, extending beyond classical multiplexing schemes \cite{Klemes2019}. Only very recently has OAM been explored in astronomy and astrophysics, as previously suggested in Ref.\ \cite{Harwit2003}. Theoretical predictions and experimental verification confirm that light passing near a rotating black hole (modeled as a Kerr black hole \cite{Kerr1963}) acquires a twist in its spatial phase distribution due to the rotating spacetime, enabling the extraction of information about the black hole’s rotation \citerange{Tamburini2011}{Tamburini2021}. Moreover, exploiting the analogy between gravitational fields and optical media within the general relativistic geometric optics framework \cite{Plebanski1960,deFelice1971}, simple formulas have been proposed to characterize the OAM of radiation from a rotating compact object \cite{Tamburini2021pra}. This allows us to construct a refined criterion for modeling Kerr-like geometries and, as we show here, to derive an explicit bound on the electric charge of M87*.

The paper is organized as follows. In Sec.\ 1, some fundamental concepts regarding SAM and OAM are reviewed; in particular, the so-called spin-orbit coupling process, which occurs when light traverses anisotropic and inhomogeneous media, is introduced. In Sec.\ 2, it is shown that this same process also takes place when light passes near rotating compact objects, ultimately leading to a simple formula for calculating OAM. In Sec.\ 3, starting from this formula, a constraint on the electric charge of M87* is derived. Finally, Sec.\ 4 is devoted to concluding remarks.

\section{SAM-to-OAM conversion in inhomogeneous media}
\label{sec:SAM-OAM optics}
As already mentioned in Sec.\ I, light carries angular momentum in two distinct forms: SAM and OAM. SAM is associated with the polarization state of the light beam, with circularly polarized photons carrying an angular momentum of $\pm \hbar$ per photon, depending on the handedness of the polarization. In contrast, OAM arises from the spatial phase distribution of the beam, where an optical field with an azimuthal phase dependence of the form \cite{Marrucci2006}
\begin{equation}
    \mathbf{E}(r, \varphi) = \mathbf{E}_{0}(r) e^{i m \varphi}
\end{equation}
carries an OAM of $m \hbar$ per photon. In the above equation, $r, \varphi$ are polar coordinates in the $xy$ plane, while $m$ is an integer representing the topological charge, which determines the helical nature of the wavefront.  A beam with $m \ne 0$ exhibits a vortex structure with a phase singularity at the center, leading to a characteristic optical vortex, see Fig.\ 1 in Ref.\ \cite{Marrucci2006}.

In homogeneous and isotropic media, SAM and OAM are independently conserved quantities. However, when light interacts with inhomogeneous and anisotropic media, these two forms of angular momentum couple. More specifically, this coupling process, also known as spin-orbit coupling, enables the conversion of SAM into OAM and occurs when a circularly polarized input beam traverses an inhomogeneous medium. 

A specific example of such a medium is the so-called $q$-plate, first introduced in Ref.\ \cite{Marrucci2006}, where the optical axis orientation varies azimuthally according to 
\begin{equation}
    \alpha(r, \varphi) = q \varphi + \alpha_{0},
\end{equation}
where $q$ and $\alpha$ are constants. The effect of the $q$-plate on the polarization state of the beam can be described using the Jones matrix formalism. In particular, the Jones matrix $\mathbf{M}$ to be applied to the field at each point of the $q$-plate transverse plane $xy$ is given by \cite{Marrucci2006}
\begin{equation}
    \mathbf{M} = \begin{bmatrix}
    \cos2\alpha(r,\varphi) & \sin2\alpha(r,\varphi) \\
    \sin2\alpha(r,\varphi) & -\cos2\alpha(r,\varphi)
    \end{bmatrix}.
\end{equation}
An incident left-circularly polarized beam described by a Jones electric field vector $\mathbf{E}_{\text{in}} = E_{0} \times [1, i]^{T}$, emerging from the medium, transforms as \cite{Marrucci2006}
\begin{equation} \label{EFtransformation}
    \mathbf{E}_{\text{out}} = \mathbf{M} \cdot \mathbf{E}_{\text{in}} = E_0 e^{i 2 q \varphi} e^{i 2\alpha_0} \begin{bmatrix}
    1 \\
    -i 
    \end{bmatrix},
\end{equation}
i.e., it becomes right-circularly polarized with an additional azimuthal phase factor $e^{i 2 q \varphi}$, indicating that the beam has acquired an OAM of $2 q \hbar$ per photon.  Conversely, a right-circularly polarized input would generate a left-circularly polarized output with an opposite OAM of $-2q\hbar$, demonstrating that the input polarization determines the sign of the generated OAM. For a schematic representation of the optical effect of a $q$-plate, see, e.g., Fig.\ 2 in Ref.\ \cite{Marrucci2013}.

In the next section, by exploiting the analogy between gravitational fields and optical media within the framework of general relativistic geometric optics, we will understand how the concepts introduced here can help us obtain a constraint on the electric charge of M87.

\section{Interplay between SAM and OAM in rotating black hole spacetimes}
\label{sec:SAM-OAM curved spacetime}

To understand the evolution of angular momentum for a light beam as it propagates through the curved spacetime surrounding a rotating black hole, one can apply the principles of geometric optics within the framework of general relativity. This approach allows to draw an analogy where light behaves as if it is traversing an optical anisotropic and inhomogeneous medium while moving through different regions of spacetime. Through this analogy, one can explore how the SAM and OAM evolve during the propagation of light.

When light is emitted from an accretion disk around a rotating black hole, it exhibits specific patterns in its OAM spectrum. These patterns are shaped by the characteristics of the source — in the case we consider here, the accretion disk itself — as well as general relativistic effects, including spacetime dragging and gravitational lensing, which can distort and rotate the light’s wavefront as it travels near the black hole. 

In this section, we will briefly review how to characterize the SAM-to-OAM conversion in the context of rotating spacetimes. Analytically, as discussed in Ref.\ \cite{Tamburini2021pra}, there exist various approaches to study this phenomenon. Here, we will focus on a particular approach which, despite its simplicity, yields interesting results. 

The idea is to simply consider the black hole as analogous to a q-plate: a left- (or right-) circularly polarized light beam, traversing the rotating spacetime, is transformed into a right- (or left-) circularly polarized beam, with an additional OAM component, in perfect analogy with what happens in the case of anisotropic and inhomogeneous media, see Sec.\ \ref{sec:SAM-OAM optics}. 

In order to see this explicitly, let us consider the Kerr metric in Boyer-Lindquist coordinates $(t,r,\vartheta,\varphi)$; the line element can be written as \cite{Ishihara1988}
\begin{multline}\label{Kerr}
    ds^2 = \frac{\sin^2 \vartheta}{\rho^2}\left[a \, dt - \left(r + a\right)^2 d\varphi\right]^2  + \frac{\rho^2}{\Delta}dr^2 \\
    + \rho^2 d\vartheta^2 - \frac{\Delta}{\rho^2} (dt - a \, \sin^2 \vartheta \, d\varphi)^2, 
\end{multline}
where the quantities $\rho^2$ and $\Delta$ are defined by
\begin{equation}
    \rho^2 \coloneqq r^2 + a^2 \cos^2 \vartheta, \quad
    \Delta \coloneqq r^2 + a^2 - 2Mr.
\end{equation}
Above, $M$ denotes the ADM mass of the black hole, while $a$ represents its angular momentum per unit mass.

The scenario we consider is a typical gravitational lensing setup, where photons emitted from a source travel close to the black hole before continuing on to reach the observer. Although we are ultimately interested in a setting where the light originates from the accretion disk surrounding the black hole, we will assume the source is ``far enough''; this assumption will turn out to be a reasonable approximation since the emission from the accretion disk is averaged across its entire extent.

By introducing reference frames at the source and observer, one can derive an equation that describes how the two components of the polarization vector, which are transverse to the beam’s propagation direction, transform between the source and the observer, namely \cite{Tamburini2021pra,Ishihara1988}
\begin{equation} \label{transformationOS}
    \begin{bmatrix}
    f_{\parallel} \\
    f_{\perp}
    \end{bmatrix}_{\text{observer}} = \begin{bmatrix}
    \cos\xi & -\sin\xi \\
    \sin\xi & \cos\xi
    \end{bmatrix}
    \begin{bmatrix}
    f_{\parallel} \\
    f_{\perp}
    \end{bmatrix}_{\text{source}}.
\end{equation}
In the expression above, the components $f_{\parallel}$ and $f_{\perp}$ represent the orientation of the electric field vector in the transverse plane, while the rotation angle $\xi$ — commonly referred to as the gravitational Faraday rotation angle — is given to linear order in the spin parameter $a$ by \cite{Tamburini2021pra,Ishihara1988}
\begin{equation} \label{rotationangle}
    \xi \simeq \frac{5\pi}{4}\frac{a M^2 \cos \vartheta_{0}}{r_{\text{min}}^{3}},
\end{equation}
where $\vartheta_{0}$ denotes the inclination angle of the black hole’s spin axis relative to the observer's line of sight, and $r_{\text{min}}$ is the minimum radial coordinate that the light beam reaches as it passes by the black hole. As can be deduced from Eq.\ \eqref{transformationOS}, a left-circularly polarized light beam (with polarization components $f_{\perp} = 1$ and $f_{\parallel} = i$) passing near the rotating gravitational lens is transformed into a right-circularly polarized beam, with an additional phase factor \cite{Tamburini2021pra,Ishihara1988}:
\begin{equation} \label{SAMOAMcp}
    \begin{bmatrix}
    \cos\xi & -\sin\xi \\
    \sin\xi & \cos\xi
    \end{bmatrix} \begin{bmatrix}
    1 \\
    i
    \end{bmatrix}_{\text{source}} = e^{-i\xi} \begin{bmatrix}
    1 \\
    -i
    \end{bmatrix}_{\text{observer}}.
\end{equation}
Now, assuming the black hole acts as a q-plate and comparing Eq.\ \eqref{SAMOAMcp} with Eq.\ \eqref{EFtransformation} of Sec.\ \ref{sec:SAM-OAM optics}, the OAM content in the radiation reaching the observer can be identified as \cite{Tamburini2021pra}
\begin{equation} \label{oam}
    \langle m \rangle \approx \frac{5\pi}{4}\frac{a M^{2} \cos \vartheta_{0}}{\Bar{r}^{3}}.
\end{equation}

A few comments are in order here:
\begin{enumerate}[(i)]
    \item The equation presented above arises from the application of the analogy between rotating geometries and inhomogeneous and anisotropic media. Specifically, similar to how optical systems, such as spiral phase plates, impart a defined OAM to a light beam (as discussed in Section \ref{sec:SAM-OAM optics}), the gravitational field of a rotating black hole induces a corresponding effect. As the light propagates through the curved spacetime around the rotating black hole, different parts of the wavefront experience varying gravitational effects, leading to path-dependent phase shifts. Although $\xi$ does not explicitly depend on the azimuthal angle, the cumulative phase shift experienced as the light interacts with the rotating black hole creates an effective azimuthal phase variation, resulting in the light
    acquiring OAM.
    \item A light beam typically consists of multiple OAM modes, each associated with a distinct $m$ value. The OAM content can be characterized by the spiral spectrum, which is a histogram of the weights of each OAM component present in the received beam (see, e.g., the lower panel of Fig.\ 3 in Ref.\ \cite{Tamburini2020}). The overall OAM is expressed as $\langle m \rangle$, reflecting the average based on the distribution of these modes. In the case at hand, since the OAM distribution is dominated by the modes $m = 0$ and $m = 1$ \cite{Tamburini2020}, the average OAM can be estimated by considering the ratio between the weights of these two components \cite{Tamburini2021}:
    \begin{equation}
        \langle m \rangle \approx \frac{\text{weight}(m = 0)}{\text{weight}(m = 1)}.
    \end{equation}
    \item Since the left-hand side of Eq.\ \eqref{oam} involves an average OAM, the radial coordinate on the right-hand side should also be interpreted as an average or effective radius; for simplicity, we renamed $r_{\text{min}}$ as $\Bar{r}$. This effective radius can be understood as the average radial coordinate at which most of the OAM observed in the emitted radiation is generated, thus characterizing the accretion disk. We refer to this average radius as the \textit{OAM emission fiducial radius}.
\end{enumerate}

The results of this section can also be extended to the Kerr-Newman black hole metric, which describes the spacetime geometry around an electrically charged and rotating compact object. In Boyer-Lindquist coordinates, the line element is given by Eq.\ \eqref{Kerr}, where the function $\Delta$ now reads \cite{Hájiček2008}
\begin{equation}
    \Delta \coloneqq r^2 + a^2 - 2Mr + \mathcal{Q}^{2}.
\end{equation}
In Ref.\ \cite{Guo2023}, a generalization of Eq.\ \eqref{rotationangle} to the case of an electrically charged and rotating Taub-Newman-Unti-Tamburino spacetime \cite{Griffiths1973} was obtained. By setting the charge $l$ to zero, one immediately finds the result for the rotation of the polarization angle of an electromagnetic wave propagating in a rotating gravitational field. Starting from this result and repeating the analysis in Ref.\ \cite{Tamburini2021pra}, one obtains a generalization of Eq.\ \eqref{oam} to the case of a charged rotating black hole, expressed as
\begin{equation} \label{averageOAMQ}
    \langle m \rangle_{\mathcal{Q}} = \frac{5\pi}{4}\frac{a M^2 \cos \vartheta_{0}}{\Bar{r}^{3}} \left(1 + \frac{\mathcal{Q}^2}{5 M^2}\right).
\end{equation}
As expected, the above expression reduces to Eq.\ \eqref{oam} if we set $\mathcal{Q} = 0$. The behavior of the average OAM, given by Eq.\ \eqref{averageOAMQ}, is plotted in Fig.\ 1 as a function of $\Bar{r}$, for different values of the electric charge $\mathcal{Q}$ and spin parameter $a$.

Assuming that M87* is described by the Kerr-Newman solution, in the next section, we will use Eq.\ \eqref{averageOAMQ} to constrain its electric charge.
\begin{figure}[t!]
\includegraphics[width=8.65cm]{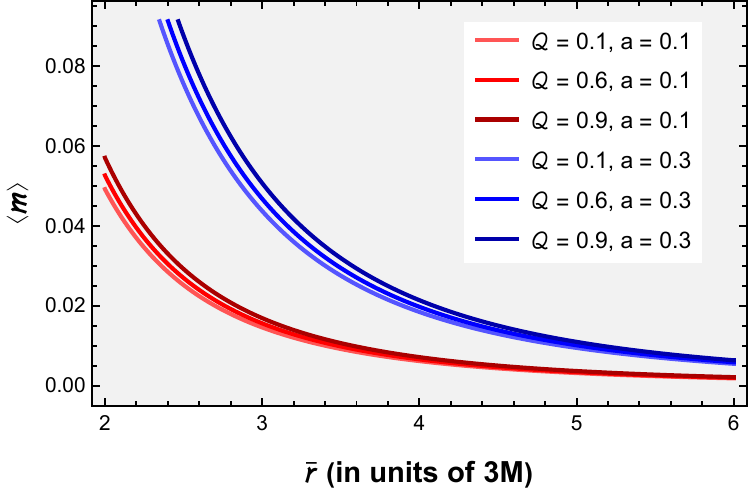}
\centering
\caption{Average OAM ($\langle m \rangle$) of radiation emitted near a charged rotating black hole as a function of the radial coordinate $r$ measured from the black hole center (in units of $3M$). Different curves correspond to varying electric charges and spin parameters. The black hole mass is normalized to $M = 1$ and the observer's viewing angle is fixed at $\vartheta_{0} = 0$, aligned with the black hole’s spin axis (polar viewing angle). The graph illustrates how the OAM content of the emitted light changes with both charge and spin, highlighting its sensitivity to these parameters. Three key features emerge clearly: (i) the OAM increases as the spin parameter $a$ increases, (ii) the OAM increases as the radial coordinate $\Bar{r}$ decreases (approaching the black hole), and (iii) the OAM increases as the electric charge $\mathcal{Q}$ increases.}
\end{figure}

\section{Charge constraint from the OAM content of M87*}
\label{sec:constraint}

In Ref.\ \cite{Tamburini2020}, the OAM content in the radiation from M87* was extracted from EHT data, assuming the black hole is described by the Kerr solution. The average OAM content has an associated relative uncertainty of approximately 3\% (at $1 \sigma$ confidence level) \cite{Tamburini2020}, so any correction to it due to charge must remain within this limit to ensure consistency. Therefore, we impose the condition
\begin{equation}
    \left | \frac{\langle m \rangle_{\mathcal{Q}} - \langle m \rangle_{\mathcal{Q} = 0}}{\langle m \rangle_{\mathcal{Q} = 0}} \right | \le \frac{\delta \langle m \rangle_{\mathcal{Q} = 0}}{\langle m \rangle_{\mathcal{Q} = 0}} \approx 0.03.
\end{equation}
By using Eqs.\ \eqref{oam} and \eqref{averageOAMQ}, and assuming $\mathcal{Q}$ to be non-negative, the inequality becomes
\begin{equation}
    \frac{\mathcal{Q}}{M} \le \sqrt{5 \frac{\delta \langle m \rangle_{\mathcal{Q} = 0}}{\langle m \rangle_{\mathcal{Q} = 0}}} \approx 0.39.
\end{equation}
Although the constraint on the electric charge we have derived from EHT data is stricter than those obtained by other methods (see, e.g., Refs.\ \cite{Kocherlakota2021,Tsukamoto2024}), it is based on the simplifying assumption that the black hole behaves as a q-plate. While this assumption is reasonable and supported by previous studies (in particular, see Ref.\ \cite{Tamburini2021pra} where Eq.\ \eqref{oam} is employed to estimate what we have defined as the OAM average fiducial radius, confirming that the OAM is, on average, generated outside of the black hole's shadow), a numerical analysis  similar to the one carried out in Ref.\ \cite{Tamburini2022} would be interesting to further confirm and refine the constraint we have found. Moreover, the formula used to obtain this constraint is linear in the spin, and higher-order corrections would be required for improved precision.

It is important to emphasize that the average OAM used in our analysis represents a well-defined observational quantity, not a model-dependent inference. Its extraction from EHT data relies on robust and tested phase-retrieval techniques, as detailed in Ref.\ \cite{Tamburini2020}.

\section{Discussion and conclusions}
\label{sec:concl}

In this work, we explored a novel method for constraining the electric charge of the supermassive black hole M87* by analyzing the OAM content in its emitted radiation. Leveraging the analogy between the gravitational field of a rotating black hole and an inhomogeneous optical medium, we interpreted the Kerr-Newman geometry as an effective q-plate that induces a SAM-to-OAM conversion in the propagating light.

By applying a simplified analytical framework previously developed in the literature \cite{Tamburini2021pra}, we derived a compact formula for the observed average OAM in terms of the black hole’s charge-to-mass ratio. Utilizing recent observational results from the EHT and the corresponding OAM measurements extracted from the twisted light of M87*, we established an upper bound on the charge-to-mass ratio of the black hole, obtaining $\mathcal{Q}/M \lesssim 0.39$. This constraint is more stringent than those derived from shadow-size analyses, although it relies on idealized assumptions such as the q-plate analogy and linear approximation in spin.

As noted in the introduction, we emphasize that the case of a charged black hole has been considered here solely as an illustrative example. In realistic astrophysical environments, black holes are expected to be electrically neutral because any excess charge would quickly be neutralized by infalling plasma. The methodology, however, is applicable to other types of charges or generalized classical black hole geometries, including non-singular solutions constructed via signature transitions \cite{Capozziello2024}, as well as spacetimes with non-trivial matter configurations, such as string clouds or dark matter halos \cite{Mustafa2024}, which may further influence OAM transfer.

It is important to stress that, although the bound we obtained on the electric charge of M87 is numerically larger than those inferred from theoretical considerations or other astrophysical observations (e.g., Faraday rotation \cite{Kuo2014}), our result is not in tension with these limits. Rather, our goal is to explore how existing EHT data — from which one can extrapolate the OAM content of the emitted radiation — can be used to place meaningful constraints.

While the assumptions considered in this paper are supported by earlier studies and yield promising results, further investigations incorporating numerical ray-tracing could improve the robustness and precision of our findings. In particular, numerical simulations could refine the mapping between the observed OAM spectrum and spacetime parameters, accounting for realistic disk geometries, additional light propagation effects, and moving beyond the linear approximation in the spin.

Furthermore, it would be valuable to explore how accretion dynamics in alternative theories or models inspired by quantum gravity (see, e.g., Refs.\ \cite{Mustafa2025,Naseer2024}), as well as dynamic lensing effects due to black hole motion \cite{Wang2025}, could lead to subtle corrections in the OAM spectrum.

This study highlights the potential of OAM-based analyses as a complementary tool in black hole astrophysics, paving the way for new tests of gravity in the strong-field regime using light’s twisted signatures.

\section*{Acknowledgements}
F. F. and G. L. acknowledge support from the University of Salerno. F. F. gratefully acknowledges Nordita for its support through the Nordita Visiting PhD Fellow Program.

\end{document}